\documentclass[conference, a4paper]{IEEEtran}
\IEEEoverridecommandlockouts
\usepackage{cite}
\usepackage{amsmath,amssymb,amsfonts}
\usepackage{algorithmic}
\usepackage{graphicx}
\usepackage{pgfplots}
\usepackage{tikz}
\usetikzlibrary{shapes,arrows, positioning, automata}
\usepackage{tkz-fct}
\usepackage{subfig}
\usepackage{textcomp}
\usepackage{xcolor}
\usepackage{caption}
\usepackage{hyperref}

\newcommand\copyrighttext{%
  \footnotesize \textcopyright 2020 IEEE. Personal use of this material is permitted.
  Permission from IEEE must be obtained for all other uses, in any current or future
  media, including reprinting/republishing this material for advertising or promotional
  purposes, creating new collective works, for resale or redistribution to servers or
  lists, or reuse of any copyrighted component of this work in other works.
  DOI: \href{https://doi.org/10.1109/MMSP48831.2020.9287096}{10.1109/MMSP48831.2020.9287096}}
\newcommand\copyrightnotice{%
\begin{tikzpicture}[remember picture,overlay]
\node[anchor=south,yshift=10pt] at (current page.south) {\fbox{\parbox{\dimexpr\textwidth-\fboxsep-\fboxrule\relax}{\copyrighttext}}};
\end{tikzpicture}%
}

\newlength\fwidth
\tikzset{cross/.style={cross out, draw=black, minimum size=2*(#1-\pgflinewidth), inner sep=0pt, outer sep=0pt},cross/.default={4pt}}
\graphicspath{figures/}
\begin{document}

\title{Key Point Agnostic Frequency-Selective Mesh-to-Grid Image Resampling using Spectral Weighting}

\author{\IEEEauthorblockN{Viktoria Heimann, Nils Genser, and Andr\'e Kaup}
\IEEEauthorblockA{\textit{Multimedia Communications and Signal Processing} \\
\textit{Friedrich-Alexander University}\\
Erlangen-Nuremberg, Germany\\
\{viktoria.heimann, nils.genser, andre.kaup\}@fau.de}

}

\maketitle
\copyrightnotice
\begin{abstract}
	Many applications in image processing require resampling of arbitrarily located samples onto regular grid positions. This is important in frame-rate up-conversion, super-resolution, and image warping among others. A state-of-the-art high quality model-based resampling technique is frequency-selective mesh-to-grid resampling which requires pre-estimation of key points. In this paper, we propose a new key point agnostic frequency-selective mesh-to-grid resampling that does not depend on pre-estimated key points. Hence, the number of data points that are included is reduced drastically and the run time decreases significantly. To compensate for the key points, a spectral weighting function is introduced that models the optical transfer function in order to favor low frequencies more than high ones. Thereby, resampling artefacts like ringing are supressed reliably and the resampling quality increases. On average, the new AFSMR is conceptually simpler and gains up to 1.2~dB in terms of PSNR compared to the original mesh-to-grid resampling while being approximately 14.5 times faster.  

\end{abstract}
\section{Introduction}
\label{sec:intro}
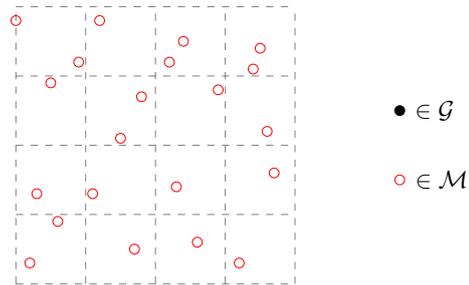
\begin{figure}
\centering
	\resizebox{.73\linewidth}{!}{
	\begin{tikzpicture}
\tkzInit[xmax=4, ymax=4]
\begin{scope}[dashed]
\tkzGrid
\end{scope}
\draw[color=red](0.2, 0.3)circle(2pt);
\draw[color=red](0.3, 1.3)circle(2pt);
\draw[color=red](0.6, 0.9)circle(2pt);
\draw[color=red](0.5, 2.9)circle(2pt);
\draw[color=red](0.9, 3.2)circle(2pt);
\draw[color=red](0, 3.8)circle(2pt);
\draw[color=red](1.2, 3.8)circle(2pt);
\draw[color=red](1.7, 0.5)circle(2pt);
\draw[color=red](1.5, 2.1)circle(2pt);
\draw[color=red](1.1, 1.3)circle(2pt);
\draw[color=red](1.8, 2.7)circle(2pt);
\draw[color=red](2.6, 0.6)circle(2pt);
\draw[color=red](2.3, 1.4)circle(2pt);
\draw[color=red](2.4, 3.5)circle(2pt);
\draw[color=red](2.9, 2.8)circle(2pt);
\draw[color=red](2.2, 3.2)circle(2pt);
\draw[color=red](3.2, 0.3)circle(2pt);
\draw[color=red](3.7, 1.6)circle(2pt);
\draw[color=red](3.5, 3.4)circle(2pt);
\draw[color=red](3.6, 2.2)circle(2pt);
\draw[color=red](3.4, 3.1)circle(2pt);

\draw[fill=black](5.5, 2.5)circle(2pt);
	\draw[color=red](5.5, 1.5)circle(2pt);
	\node at (6, 2.5) {$\in \mathcal{G}$};
	\node at (6.1, 1.5) {$\in \mathcal{M}$};
\end{tikzpicture}
}

\caption{\label{Fig:Database}{Definition of point sets. The available samples on the floating mesh $\mathcal{M}$ are denoted as red circles. Black dots represent the pixels on the regular grid $\mathcal{G}$ that are to be reconstructed. Using FSMR, these points are interpolated first using bicubic interpolation and are refered to as key points. They are disregarded using the newly proposed AFSMR. }}
\vspace{-.4cm}
\end{figure}

In image processing, the need for reconstructing samples from non-integer positions onto regular grid positions is of high importance for many applications. For example, this is required when different warped meshes are combined to one single-view point like in view generation \cite{2019_Lee_ViewGeneration}. The same problem holds for multi-camera setups and stereo matching applications \cite{2002_Kolmogorov_MultiCamera, 2007_Tarel_ICIP}. Furthermore, non-integer positions arise in scenarios that take sub-pixel accuracy implicitly as e.g. super-resolution \cite{2003_Park_SROverview} or frame-rate up-conversion \cite{2014_Lee_FRUC}. The arbitrarily located non-integer positions are referred to as \textit{mesh}, the integer positions as \textit{grid} in the following. In literature, mesh points are also known as scattered data and thus, the solution of this interpolation task is refered to as scattered data interpolation. In Fig.~\ref{Fig:Database} the mesh positions are given as red circles. The task is to resample the red circles onto regularly spaced grid positions, which are given as black dots in Fig.~\ref{Fig:Database}. They are located on the intersections of the dashed lines which aim to visualize the grid. \par 
Common state-of-the-art methods that address this problem statement are presented in the next section. In line with this, one of those state-of-the-art methods is named frequency-selective mesh-to-grid resampling (FSMR) \cite{2017_Koloda_FSMR}. To overcome the drawback of computational complexity and to enhance the resampling quality further, the novel key point agnostic FSMR (AFSMR) with spectral weighting is proposed. The new AFSMR is presented in Sec.~\ref{sec:newfsmr}. Subsequently, the techniques are evaluated in terms of quality and run time in Sec. \ref{sec:results}. The paper concludes with Sec.~\ref{sec:conclusion}.

\section{State-Of-The-Art}
\label{sec:fsmr}
In mesh to grid resampling pixels that are located on arbitrary floating mesh positions have to be resampled onto regular grid positions. Grid positions hold integer coordinates whereas floating mesh positions hold non-integer positions. We refer to this problem as image resampling or scattered data interpolation. To tackle this problem, common interpolation techniques can be applied. Using bilinear interpolation \cite{2002_Amidror_InterpolationMethodsSurvey} the information on the grid can be achieved by taking the mean of the neighboring floating mesh data points. A similar approach holds for bicubic interpolation \cite{2002_Amidror_InterpolationMethodsSurvey} where polynomials are used to fit the original data and interpolate the unknown grid information. Also Lanczos interpolation can be used to solve the interpolation task. \\
Other approaches to solve the resampling problem can be found in the context of image reconstruction and more specifically in the area of the reconstruction of irregularly sampled image data. Nevertheless, irregularly sampled data is not the same as scattered data. In most cases, if one refers to irregularly sampled data, the data is still sampled on fixed grid positions. Scattered data are usually located on arbitrary non-integer positions. However, the reconstruction of images from irregularly sampled data can handle differences in pixel density and the reconstruction from arbitrarily distributed original data. 
A very common representative of this group of reconstruction techniques is the kernel regression approach by Takeda et al. \cite{2007_Takeda}. Other reconstruction methods use Delaunay triangulation \cite{2002_Bose}, second-order wavelets \cite{2004_Bose}, or moving least squares \cite{2006_Bose} approaches to handle the irregularly sampling problem. \\
Another method that can be used for resampling and reconstruction is the model-based Frequency-Selective Mesh-to-Grid Resampling (FSMR) \cite{2017_Koloda_FSMR}. It originates from frequency-selective reconstruction \cite{2015_Seiler_ResamplingImages}, which was developed over the last years \cite{2017_Genser_FastFSR, 2018_Genser_PCS}. It is designed to solve quarter sampling problems and thus can be seen as a representative of the irregular sampling group. FSMR is a versatile technique to handle any kind of scattered data interpolation problem. Furthermore, FSMR is a powerful technique that yields excellent resampling quality \cite{2017_Koloda_FSMR}. A great drawback of FSMR however is, that a huge data base is incorporated into model generation due to the pre-estimation of key points and, thereby, the run time is quite long. As FSMR is important for the ongoing work, we will shortly summarize it in the following.\par 
As already described, arbitrarily distributed mesh pixels $\mathcal{M}$ are resampled onto regularly distributed grid coordinates $\mathcal{G}$. The grid coordinates are interpolated in a first step using bicubic interpolation to stabilize FSMR. This step yields a high increase in the number of underlying data points. We denote these points as key points $\mathcal{G}_\text{key}$. A visualization of the used (red) and the to be reconstructed coordinates (black) of FSMR is given in Fig.~\ref{Fig:Database}. To generate a model, FSMR uses the assumption that an image $f[m,n]$ can be locally represented as a weighted superposition of two-dimensional basis functions $\varphi_{(k,l)}$. Therefore, the image is partitioned into blocks with a support area around. It holds in total $M\times N$ pixels. In a given reconstruction area $\mathcal{A} = \mathcal{M} \cap \mathcal{G}_\text{key}$, the image signal is defined as
\begin{equation}
\label{Eq:image}
f[m, n] = \sum_{(k, l) \in \mathcal{K}} c_{(k, l)} \varphi_{(k, l)}[m, n], 
\end{equation}   
with expansion coefficients $c_{(k, l)}$, the set of available basis functions $\mathcal{K}$ and pixel cooordinates $(m, n) \in \mathcal{A}$. As orthogonal bases from the discrete cosine transform (DCT) are considered, the expansion coefficients can be interpreted as transform  coefficients of the inverse transform. \\
The model is generated iteratively within the reconstruction area $\mathcal{A}$. It is initialized to zero, i.e. $g^{(0)} = 0\;\;\forall (m,n)\in\mathcal{A}$.
In iteration $\nu$ the model is given according to 
\begin{equation}
\label{Eq:modelGeneration}
g^{(\nu)}[m, n] = g^{(\nu -1)}[m, n] + \hat{c}_{(u, v)} \varphi_{(u, v)}[m, n],
\end{equation}
with the chosen basis function $\varphi_{(u, v)}$ in iteration $\nu$ and the corresponding estimated expansion coefficient $\hat{c}_{(u, v)}$ where $u$ and $v$ denote the selected frequency coefficients. In order to find the best fitting basis function, the weighted residual $r$ is determined in every iteration step
\begin{equation}
\label{Eq:Residual}
r^{(\nu)} = (f[m,n] - g^{(\nu)} [m,n]) w[m,n]\,.
\end{equation}
The spatial weighting function 
\begin{equation}
\label{Eq:SpatialWeighting}
w[m, n] = \begin{cases}
\rho^{\sqrt{(m-\frac{M-1}{2})^2 + (n-\frac{N-1}{2})^2}} & \forall(m,n)\in\mathcal{M}, \\
\alpha\rho^{\sqrt{(m-\frac{M-1}{2})^2 + (n-\frac{N-1}{2})^2}} &\forall(m,n)\in\mathcal{G}_\text{key},
\end{cases}
\end{equation}
is used to weight the influence of every single pixel on the reconstruction. The isotropic window function smoothly decreases with distance to the center of the reconstruction area. Furthermore, a lower weight is assigned to key points as these are only estimations and not as trustworthy as original samples. The spatial weighting is controlled by parameter $\alpha$. As further explained in \cite{2017_Koloda_FSMR}, $\alpha$ is chosen adaptively to the density of pixels in $\mathcal{A}$. 
To obtain the expansion coefficient $\hat{c}^{(\nu)}_{(k, l)}$, the weighted residual energy $E^{(\nu)}$ can be obtained \cite{2017_Koloda_FSMR}. Following \cite{2005_Kaup_FSEOrig} and setting the derivative of $E^{(\nu)}$ with respect to $\hat{c}_{(u, v)}$ to zero, the expansion coefficient is given as 
\begin{equation}
\label{Eq:estExpCoeff}
\hat{c}^{(\nu)}_{(k, l)} = \frac{\sum_{(m,n)\in \mathcal{A}} r^{(\nu-1)}[m,n] \varphi_{(k,l)}[m,n]}{\sum_{(m,n)\in \mathcal{A}} w[m,n](\varphi_{(k, l)}[m,n])^2}.
\end{equation}
In every iteration step one basis function is chosen. For the best choice the difference of the residual energy from $E^{(\nu-1)}$ to $E^{(\nu)}$ is calculated 
\begin{equation}
\label{Eq:deltaE}
\Delta E_{(k, l)}^{(\nu)} = (\hat{c}^{(\nu)}_{(k, l)})^2 \sum_{(m,n)\in \mathcal{A}}w[m,n](\varphi_{(k, l)}[m,n])^2. 
\end{equation}
The frequency indeces of the basis function that gives the greatest reduction in residual energy are selected according to 
\begin{equation}
\label{Eq:uArgmax}
{(u, v)} = \underset{{(k, l)}}{\mathrm{argmax}} \Delta E_{(k, l)}^{(\nu)}
\end{equation}
and the corresponding basis function $\varphi_{(u,v)}$ is subsequently used in \eqref{Eq:modelGeneration}. The procedure is repeated until a stopping criterium is met. This can either be a maximum number of iterations or a maximum decrease in residual energy. If the model generation is finished, the points $(m,n) \in \mathcal{G}$ are taken out of the model and used as grid point reconstruction. This procedure is repeated until all image blocks have been reconstructed \cite{2017_Koloda_FSMR}.\\
\section{Novel key point agnostic FSMR}
\label{sec:newfsmr}
Key points are important in FSMR since they restrict the possible solutions during model generation and thereby stabilize it. As described in the last section, key points are estimations based on cubic interpolation. Thus, these points can be considered a workaround, as they restrict the reconstruction to a low-frequency model. If they are included in the model generation, the model is adapted to these suboptimal points. To overcome this drawback, we propose to disregard key points and generate the image model solely on the original floating mesh points using a frequency prior for stabilization. Thereby, the number of input points decreases drastically. As all calculations have to be conducted for all input points, the number of computations decreases considerably. Consequently, the run time of AFSMR is reduced significantly while it still generates a high-quality image model. AFSMR is presented in detail in the following.
\subsection{Key point agnostic resampling}
The same approach as in Sec. \ref{sec:fsmr} is conducted. An image signal is represented according to \eqref{Eq:image} and the corresponding model is generated iteratively as given in \eqref{Eq:modelGeneration}. Now, the key points are disregarded and only the original mesh points $\mathcal{M}$ are used. The block partitioning remains. Thus, the reconstruction area is reduced to $\mathcal{A} = \mathcal{M}$. In order to obtain the best fitting basis function in iteration $\nu$ the residual between the available original signal $f[m,n]$ and the current model is considered
\begin{equation}
\label{Eq:ResidualNew}
r^{(\nu)} = f[m,n] - g^{(\nu)}[m,n].
\end{equation}
In contrast to FSMR, the residual is not weighted in AFSMR. FSMR considers key points and thus, the residual on mesh and grid has to be treated differently. As no key points are considered in AFSMR, no weighting has to be applied. Nevertheless, the spatial weighting $w[m,n]$ function still has to be used in the computation of the weighted residual energy function $E$ for the given reconstruction area $\mathcal{A}$
\begin{equation}
E^{(\nu)} = \sum_{(m,n) \in \mathcal{A}} w[m,n]\left(r^{(\nu)}[m,n]\right)^2.
\end{equation}
The spatial weighting is defined as a decaying isotropic window, i.e., 
\begin{equation}
\label{Eq:SpatialWeightingNew}
w[m, n] = 
\rho^{\sqrt{(m-\frac{M-1}{2})^2 + (n-\frac{N-1}{2})^2}}, \;\; \forall (m,n) \in \mathcal{M}.
\end{equation}
It is solely determined on the original mesh positions. Using these changes, the expansion coefficient in iteration $\nu$ is expressed as 
\begin{equation}
\label{Eq:estExpCoeffNew}
\hat{c}^{(\nu)}_{(k, l)} = \frac{\sum_{(m,n)\in \mathcal{A}} r^{(\nu-1)}[m,n] \varphi_{(k, l)}[m,n]w[m,n]}{\sum_{(m,n)\in \mathcal{A}} w[m,n](\varphi_{(k, l)}[m,n])^2}.
\end{equation}
The selection of the best fitting basis function in the current iteration step $\nu$ can be pursued following \eqref{Eq:deltaE} and \eqref{Eq:uArgmax}. Due to the disregard of the key points, an additional spectral weighting is suggested for the selection of the best fitting basis function in order to restrict the solution space for the selection of a suitable coefficient to the most probable frequencies. It is described in further detail in the next section. The model generation is finished if one of the already described stopping criteria is met.\\
The values on the grid positions still have to be determined. Therefore, the generated model offers the estimated expansion coefficients $\hat{c}_{(k, l)}$ for all possible basis functions. These coefficients are taken and inserted into \eqref{Eq:image}. The image signal is evaluated on the grid positions  
\begin{equation}
\label{Eq:ResampledImage}
f[m, n] = \sum_{(k, l) \in \mathcal{K}} \hat{c}_{(k, l)} \varphi_{(k, l)}[m, n], \;\; \forall (m,n) \in \mathcal{G}.
\end{equation} 
Finally, the image signal $f[m, n]$ on the grid is taken while omitting the original mesh points.
\subsection{Spectral Weighting}
\begin{figure}
\setlength\fwidth{0.32\textwidth}
\centering
\input{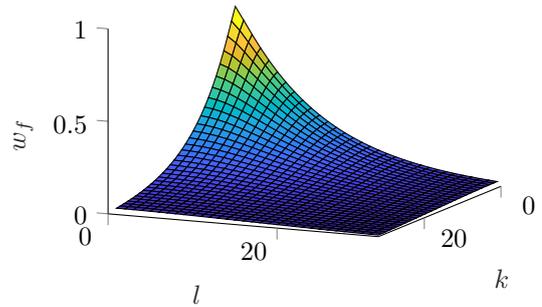}
\caption{\label{Fig:wf} Exemplary spectral weighting function $w_f$ for ${k=l=32}$ frequency indexes in horizontal and vertical direction, respectively, and ${\sigma = 0.9}$.}
\vspace{-.3cm}
\end{figure}
\begin{figure}
\centering
\resizebox{.7\linewidth}{!}{

\definecolor{colorflow}{rgb}{0.00000,0.44700,0.74100}%
\tikzstyle{decision} = [diamond, draw, fill=white!15, 
    text width=4.5em, text badly centered, node distance=3cm, inner sep=0pt]
\tikzstyle{block} = [rectangle, draw, fill=white!15, 
    text width = 20em, text centered, rounded corners, minimum height=3em]
 \tikzstyle{blockk} = [rectangle, draw, fill=colorflow, 
    text width = 20em, text centered, rounded corners, minimum height=3em]
\tikzstyle{blockkk} = [rectangle, draw=white, fill=white!15, 
    text width = 20em, text centered, rounded corners, minimum height=3em]
\tikzstyle{line} = [draw, -latex']
\tikzstyle{cloud} = [draw, ellipse,fill=white!20, node distance=5cm,
    minimum height=2em]
    
\begin{tikzpicture}[node distance = 2cm, auto]
    \node [blockkk] (init) {Original signal $f[m,n]$, $[m,n]\in \mathcal{M}$. \\ \textcolor{colorflow}{No key points at $[m,n]\in \mathcal{G_{\text{key}}}$ are estimated.}};
    \node [block, below of=init] (residual) {Calculate residual $r^{(\nu)}[m,n]$};
    \node [block, below of=residual] (expcoeff) {Compute and update estimated expansion coefficients $\hat{c}_{(k,l)}$ for every basis function};
    \node [block, below of=expcoeff] (energy) {Calculate energy decrease $\Delta E^{(\nu)}$};
    \node [block, below of=energy] (wf) {\textcolor{colorflow}{Spectral Weighting $w_f(k,l)$}};
    \node [block, below of = wf] (selection) {Selection of best fitting basis function};
    \node [decision, below of=selection, node distance = 3cm] (stop) {Stopping criterium met?};
    \node [block, below of= stop, node distance = 3cm] (grid) {Obtain signal on grid positions, i.e. final $g[m,n]$, $[m,n]\in \mathcal{G}$};
    
    \coordinate[right of=residual] (a1);  
    \coordinate[right of=stop] (e1); 
    
    \path [line] (init) -- (residual);
    \path [line] (residual) -- (expcoeff);
    \path [line] (expcoeff) -- (energy);
    \path [line] (energy) -- (wf);
    \path [line] (wf) -- (selection);
    \path [line] (selection) -- (stop);
    \path [line] (stop) -| node [near start]{No}([xshift=3.0cm]e1) -- node[sloped, anchor=center, above, text width = 6cm]{Generated model $g^{(\nu)}[m,n]$}([xshift=3.0cm]a1) -- (residual);
    \path [line] (stop) -- node[near start]{Yes} (grid);

\end{tikzpicture}
}
\caption{\label{Fig:FlowGraph}Flow graph summarizing the proposed AFSMR. The main improvements in AFSMR are highlighted in blue.}
\vspace{-.4cm}
\end{figure}
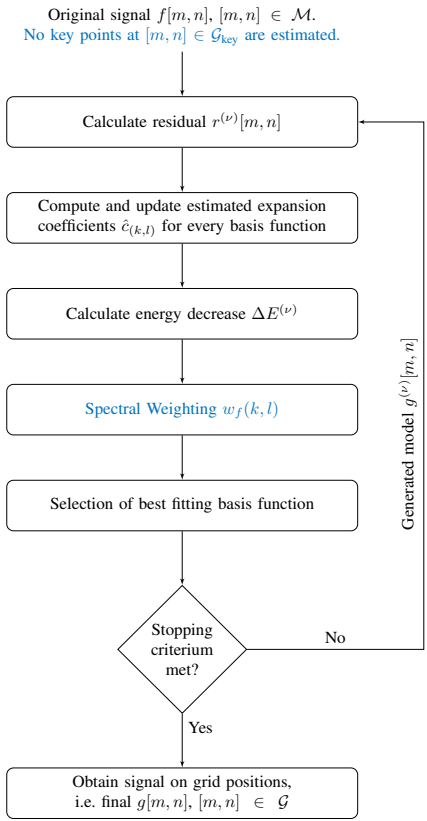

The disregard of key points leads to a smaller basis of points that are used for model generation. Hence, the solution space for the model increases. Especially high frequencies are additionally included. Supplemental high frequencies might cause artifacts like ringing and noise because natural images mainly contain low frequencies \cite{2000_Lam_DCTCoeffAnalysis}. Nevertheless, they should not be as strictly suppressed as in FSMR, which utilizes bicubic grid points. Therefore, the solution space is restricted by taking a prior frequency distribution into account. For this purpose, an additional spectral weighting is used to attenuate the high frequencies in the selection of the best fitting basis function. Inspired by the optical transfer function \cite{2015_Seiler_ResamplingImages} which attenuates high frequencies more than low ones, the spectral weighting function is expressed as
\begin{equation}
\label{Eq:FreqWeighting}
w_{f}(k,l) = \sigma^{\sqrt{k^2 + l^2}},
\end{equation}
where $\sigma \in [0,1]$ parametrizes the decay and $k$, $l$ denote the horizontal and vertical frequency indexes, respectively. The isotropic decaying window smoothly decreases with increasing frequency. It favors low frequencies in order to avoid ringing artifacts. Nevertheless, if high frequencies are dominant they are still allowed to be included in the model generation in order to resample fine structures. The spectral weighting function is visualized in Fig.~\ref{Fig:wf}. Using the spectral weighting, the selection process results in
\begin{equation}
{(u, v)} = \underset{{(k, l)}}{\mathrm{argmax}} \left( \Delta E_{(k, l)}^{(\nu)} w_{f}(k,l) \right),
\end{equation}
where $u$ and $v$ denote the selected frequency coefficients. Hence, low frequency basis functions are preferably selected, but high frequency coefficients are still allowed in model generation. \\
A summary of the whole newly proposed key-point agnostic resampling with spectral weighting is shown in the flow chart in Fig.~\ref{Fig:FlowGraph}. The improvements for AFSMR in contrast to FSMR are highlighted in blue.

\section{Experimental Results}
\label{sec:results}
In this section, experiments, parameter settings and evaluation datasets are presented first. Then, the results in terms of quality and run time are shown and analyzed.
\subsection{Experiments and Settings}
In order to thoroughly test the performance of AFSMR, a series of tests on natural images is conducted. A synthetic testing framework is employed in order to evaluate the quality in terms of PSNR and SSIM using ground truth data. Therefore, sets of mutually canceling affine transforms are considered as described in \cite{2005_Vazquez_ReconstructionNonUniformImg}. The affine transformation is defined as following:
\begin{equation}
	\boldsymbol{y = Tx + t},
\end{equation}
where $\boldsymbol{y}\in \mathbb{R}^2$ are the resulting coordinates. The matrix $\boldsymbol{T}$ is a $2\times2$ invertible transform matrix, $\boldsymbol{x}\in\mathbb{R}^2$ are the original positions and $\boldsymbol{t} \in \mathbb{R}^2$ is a translation vector. The following sequences of consecutive affine transforms are used: 
\begin{itemize}
	\item \textit{Rotation}: A forth and back rotation of angles from $10$ to $40$ degrees in steps of $2.5$ degrees,
	\item \textit{Zoom 15\%}: A 15\% zoom-in and zoom-out given by the sequence \{$\boldsymbol{T}_Z$, $\boldsymbol{T}_Z^{-1}$\} with $\boldsymbol{T}_Z = \begin{bmatrix}
	1.15 & 0\\
	0 & 1.15
	\end{bmatrix},$
	\item \textit{Affine}: A sequence of four affine transforms given by \{$\boldsymbol{T}_2^{-1}$, $\boldsymbol{T}_1$, $\boldsymbol{T}_1^{-1}$, $\boldsymbol{T}_2$\} with $\boldsymbol{T}_1 = \begin{bmatrix}
	1.2 & 0.1954\\
	0 & 1
	\end{bmatrix}$ and
	$\boldsymbol{T}_2 = \begin{bmatrix}
	1 & 0\\
	0.1954 & 1.2
	\end{bmatrix}.$
\end{itemize}
For the model-based experiments, block sizes of ${8\times8}$ pixels are used. Additionally, a support area of $8$ pixels width around the block is considered. Hence, at one time ${M\times N=24\times24}$~pixels are used for model generation. Furthermore, $1000$ iterations are taken. The spatial weighting function uses ${\rho=0.8}$. The spectral weighting function uses a basis of ${\sigma=0.9}$. 
The parameteres are trained on the set of KODAK images \cite{KODAK}. Subsequently, the tests are pursued on the TECNICK dataset \cite{TECNICK}. It provides 100 images of real-world data. The images show a resolution of ${1200\times 1200}$ pixels. All images used are $8$~bit grayscale images.
\subsection{Evaluation}
\begin{table}[t!]
	\centering
	\caption{{\label{Tab:ResultsQuality}Average PSNR in dB for TECNICK \cite{TECNICK} dataset. Best numbers are given in bold face.}}
	\begin{tabular}{|l|c|c|c|c|}
		\hline
		& Zoom 15\% & Rot $15^\circ$ & Rot $30$\textdegree & Affine \\
		\hline
		Linear & 39.0 & 37.7 & 37.8 & 35.3 \\
		\hline
		Cubic & 39.0 & 37.9 & 38.1 & 33.5\\
		\hline
		Lanczos & 37.7 & 40.6 & 41.2 & 35.3\\
		\hline 
		Kernel Reg. & 29.3 & 29.0 & 29.2 & 24.2 \\
		\hline
		MLS & 12.0 & 12.2 & 12.4 & 8.3\\
		\hline
		FSMR & 47.9 & 47.3 & 44.8 & 42.2\\
		\hline
		FSMR w/o key points & 47.4 & 47.5 & 45.1 & \textbf{42.8}\\
		\hline
		AFSMR & \textbf{49.1} & \textbf{47.9} & \textbf{47.2} & \textbf{42.8} \\
		\hline
	\end{tabular}
\end{table}

\begin{table}[t!]
	\centering
	\caption{{\label{Tab:ResultsQualitySSIM}Average SSIM for TECNICK \cite{TECNICK} dataset. Best numbers are given in bold face.}}
	\begin{tabular}{|l|c|c|c|c|}
		\hline
		& Zoom 15\% & Rot $15^\circ$ & Rot $30$\textdegree & Affine \\
		\hline
		Linear & 0.9833 & 0.9764 & 0.9764 & 0.9660 \\
		\hline
		Cubic & 0.9899 & 0.9812 & 0.9830 & 0.9675\\
		\hline
		Lanczos & 0.9806 & 0.9874 & 0.9903 & 0.9736\\
		\hline 
		Kernel Reg. & 0.8998 & 0.8944 & 0.8994 & 0.7471 \\
		\hline
		MLS & 0.0912 & 0.1000 & 0.1168 & 0.0875\\
		\hline
		FSMR & 0.9965 & 0.9949 & 0.9921 & 0.9878\\
		\hline
		FSMR w/o key points & 0.9950 & 0.9945 & 0.9921 & 0.9883 \\
		\hline
		AFSMR & \textbf{0.9969} & \textbf{0.9954} & \textbf{0.9948} & \textbf{0.9888} \\
		\hline
	\end{tabular}
\vspace{-.2cm}
\end{table}
The performance of the novel AFSMR is compared to other state-of-the-art reconstruction methods. First of all, FSMR is used as it was shown that it is the best performing reconstruction algorithm so far \cite{2017_Koloda_FSMR}. Second, we show results for FSMR w/o key points which is equivalent to AFSMR if no spectral weighting is applied. Furthermore, bilinear \cite{2002_Amidror_InterpolationMethodsSurvey} and bicubic~\cite{2002_Amidror_InterpolationMethodsSurvey} interpolation are used as commonly used interpolation methods. Additionally, Lanczos interpolation \cite{2002_GonzalezWoods_DigitalImageProcessing} is applied. These are very common interpolation methods. Moreover, the experiments are pursued on two techniques which were mainly developed for the reconstruction of irregularly sampled data. The classical kernel regression method (Kernel Reg.) by Takeda et al. \cite{2007_Takeda} and the moving least squares approach (MLS) by Bose et al. \cite{2006_Bose} are incorporated. 
\subsubsection{Quality}
\begin{figure}[t!]
	\setlength\fwidth{0.33\textwidth}
%
\definecolor{mycolor1}{rgb}{0.00000,0.44700,0.74100}%
\definecolor{mycolor2}{rgb}{0.85000,0.32500,0.09800}%
\definecolor{mycolor3}{rgb}{0.92900,0.69400,0.12500}%
\definecolor{mycolor4}{rgb}{0.49400,0.18400,0.55600}%
\definecolor{mycolor5}{rgb}{0.46600,0.67400,0.18800}%
\definecolor{mycolor6}{rgb}{0.26600,0.87400,0.8800}%
\definecolor{mycolor7}{rgb}{0.1, 0.2, 0.9}%
\definecolor{mycolor8}{rgb}{0, 0, 0}%
\begin{tikzpicture}

\begin{axis}[%
width=0.993\fwidth,
height=0.577\fwidth,
at={(0\fwidth,0\fwidth)},
scale only axis,
xmin=10,
xmax=40,
xlabel style={at={(0.5,0.04)}, font=\color{white!15!black}},
xlabel={angle in deg},
ymin=10,
ymax=49,
ylabel style={at={(0.06,0.44)}, font=\color{white!15!black}},
ylabel={PSNR in dB},
axis background/.style={fill=white},
legend style={at={(0.5,1.03)}, anchor=south, legend columns=3, legend cell align=left, align=left, draw=white!15!black}
]
\addplot [line width = 0.5mm, color=mycolor8]
  table[row sep=crcr]{%
10	47.8492883111677\\
12.5	47.6430659782357\\
15	47.4960625005617\\
17.5	47.3966313789534\\
20	47.310122206471\\
22.5	47.1950661632598\\
25	47.0562746456983\\
27.5	46.9472802889749\\
30	46.888832236862\\
32.5	46.8758527935947\\
35	46.8509866743365\\
37.5	46.8299042375579\\
40	46.7988189407448\\
};
\addlegendentry{\small AFSMR}

\addplot [line width = 0.5mm, color=mycolor1]
  table[row sep=crcr]{%
10	48.3432\\
12.5	47.7373\\
15	47.3122\\
17.5	46.9942\\
20	46.7443\\
22.5	46.3928\\
25	45.8287\\
27.5	45.2671\\
30	45.0926\\
32.5	44.9866\\
35	44.6365\\
37.5	43.9995\\
40	43.4626\\
};
\addlegendentry{\small FSMR w/o key points}

\addplot [line width = 0.5mm, dashed, color=mycolor2]
  table[row sep=crcr]{%
10	47.9789961217132\\
12.5	47.5761017755693\\
15	47.3054072893681\\
17.5	47.2063924915021\\
20	47.1804352694414\\
22.5	46.9867863604768\\
25	46.4612495561276\\
27.5	45.5760675324659\\
30	44.7595605097895\\
32.5	44.0769984653223\\
35	43.0221620458078\\
37.5	41.6635573279878\\
40	40.7331765413981\\
};
\addlegendentry{\small FSMR}

\addplot [line width = 0.5mm, color=mycolor3]
  table[row sep=crcr]{%
10	40.8609107814134\\
12.5	40.7314456291621\\
15	40.5497612499655\\
17.5	40.7871221893372\\
20	40.8859372763407\\
22.5	40.9185862726228\\
25	41.1153286029234\\
27.5	40.8237437748148\\
30	41.1704338034782\\
32.5	41.2628248045219\\
35	41.2464682671196\\
37.5	41.2990056551848\\
40	41.3031643956446\\
};
\addlegendentry{\small Lanczos}

\addplot [line width = 0.5mm, color=mycolor4, thick, dotted]
  table[row sep=crcr]{%
10	37.7750627975806\\
12.5	37.6708075390426\\
15	37.8696097295116\\
17.5	38.4754370465322\\
20	38.683295422381\\
22.5	38.2273344412627\\
25	38.386831791874\\
27.5	38.5910198308547\\
30	38.0809789338844\\
32.5	38.8686096406564\\
35	39.2424672415283\\
37.5	38.7375594948759\\
40	39.8330560499971\\
};
\addlegendentry{\small Cubic}

\addplot [line width = 0.5mm, dotted, color=mycolor5, mark=*, mark options={scale=0.5}]
  table[row sep=crcr]{%
10	37.6845001361114\\
12.5	37.7093633988922\\
15	37.7266673947317\\
17.5	37.7189932893028\\
20	37.7631371345804\\
22.5	37.7767420650498\\
25	37.7793538664171\\
27.5	37.7915172274412\\
30	37.7855425977216\\
32.5	37.8057306324447\\
35	37.8057747023064\\
37.5	37.8095872101185\\
40	37.802918160577\\
};
\addlegendentry{\small Linear}

\addplot [line width = 0.5mm, color=mycolor6]
  table[row sep=crcr]{%
10	28.9825392675252\\
12.5	28.9945985182451\\
15	28.9992663828348\\
17.5	29.006906655217\\
20	29.0266287939086\\
22.5	29.0891187868822\\
25	29.1443418350597\\
27.5	29.1939085895631\\
30	29.2352528742048\\
32.5	29.280465839503\\
35	29.335307453626\\
37.5	29.3763698766454\\
40	29.4064914191272\\
};
\addlegendentry{\small Kernel Reg.}

\addplot [dotted, thick, color=mycolor7]
  table[row sep=crcr]{%
10	12.3709976733013\\
12.5	12.4208979097332\\
15	12.1804751009791\\
17.5	12.0715316801138\\
20	12.0514776617806\\
22.5	12.1547294870146\\
25	12.4822471173758\\
27.5	12.5036659846251\\
30	12.3701874689316\\
32.5	12.1317758930068\\
35	12.0171856416077\\
37.5	12.3415578330382\\
40	12.8057007766079\\
};
\addlegendentry{\small MLS}

\end{axis}
\end{tikzpicture}%
	\caption{{\label{Fig:ResultsQualityRotation} Average PSNR in dB for TECNICK \cite{TECNICK} dataset for \textit{Rotation} of angles from 10 to 40 degrees.}}
	\vspace{-.2cm}
\end{figure}
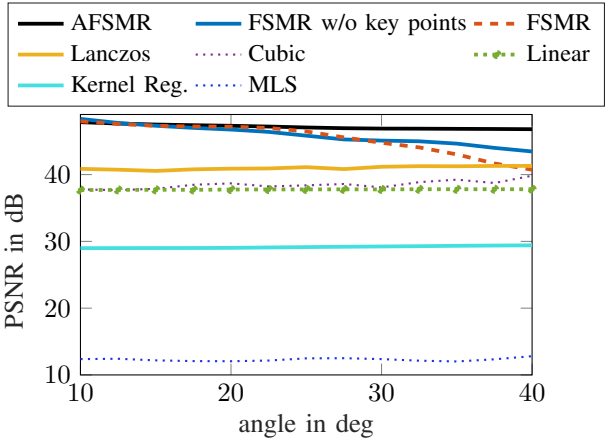

\begin{figure}[t!]
	\centering
	\subfloat[][]{
		\label{Fig:Results_Visual_Orig}
	\includegraphics[width=0.41\columnwidth]{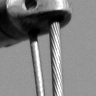}
	}
\,
		\subfloat[][]{
		\label{Fig:Results_Visual_Lin}
		\includegraphics[width=0.41\columnwidth]{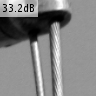}
	}
\,
\subfloat[][]{
	\label{Fig:Results_Visual_Cub}
	\includegraphics[width=0.41\columnwidth]{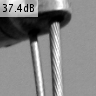}
}
\,
\subfloat[][]{
	\label{Fig:Results_Visual_Lanczos}
	\includegraphics[width=0.41\columnwidth]{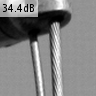}
}
\,
\subfloat[][]{
	\label{Fig:Results_Visual_Kernel}
	\includegraphics[width=0.41\columnwidth]{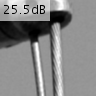}
}
\,
\subfloat[][]{
	\label{Fig:Results_Visual_Bose}
	\includegraphics[width=0.41\columnwidth]{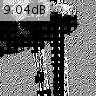}
}
\,
\subfloat[][]{
	\label{Fig:Results_Visual_Old}
	\includegraphics[width=0.41\columnwidth]{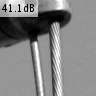}
}
\,
\subfloat[][]{
	\label{Fig:Results_Visual_New}
	\includegraphics[width=0.41\columnwidth]{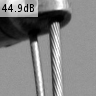}
}
\vspace{-.2cm}
\caption{{\label{Fig:Results_Visual} An extract of Image 92 of TECNICK dataset for \textit{Zoom 15\%} is given. In (a) the original is shown. The used reconstruction method is (b) bilinear, (c) bicubic, (d) Lanczos, (e) Kernel Regression, (f) MLS, (g) FSMR, and (h) AFSMR. PSNR is indicated for the respective extract. Best be viewed enlarged on a screen. }}
\vspace{-.5cm}
\end{figure}

The quality is measured in terms of PSNR and SSIM. As some state-of-the-art models yield inaccuracies in the border regions, the outer border of 24 pixels  width is neglected for all interpolation methods in both cases for the evaluation. The results for \textit{Affine}, \textit{Rotation} of 15 and 30 degrees, and  \textit{Zoom 15\%} are given explicitly in Tab.~\ref{Tab:ResultsQuality} in terms of PSNR and in Tab.~\ref{Tab:ResultsQualitySSIM} in terms of SSIM.\newline
The consecutive application of four affine transforms in the \textit{Affine} case shows that AFSMR performs best in terms of PSNR and SSIM. A gain of $0.6$~dB regarding PSNR and 0.0020 regarding SSIM can be achieved for AFSMR relative to FSMR. Furthermore, the tables show that for the two rotation cases, the novel AFSMR performs best. Additionally, it can be seen that the gain of AFSMR over FSMR increases with the angle. This behavior is also shown in Fig.~\ref{Fig:ResultsQualityRotation}. Here, the results for \textit{Rotation} are given in terms of PSNR. In this figure, the results are shown for the angles from 10 to 40 degrees in steps of $2.5$ degrees. Obviously, linear, cubic, and Lanczos interpolation perform worse than the two model-based techniques but their performance is approximately constant. The kernel regression method and MLS are outperformed by all other shown methods for all shown test cases. The reason for the rather bad performance of the kernel regression approach is that this method is originally designed for the reconstruction of irregularly sampled data, i.e. the technique is mainly focused on input data which is lying on integer grid positions. The same holds for the MLS approach. Additionally, the inverse of the basis functions multiplied with the spatial weighting function \cite{2006_Bose} leads in many cases to infinity because of the fast decay to zero of the spatial weighting function. Hence, a proper calculation of every pixel cannot be guaranteed for MLS. Nevertheless, we show all results for MLS as well. In the \textit{Rotation} case, the quality of FSMR drops drastically by nearly $8$~dB with increasing angle. If FSMR is used without key points, the drop shrinks to approximately $5$~dB. If the proposed spectral weighting is used additionally, i.e. AFSMR is applied, the drop is reduced to approximately $1$~dB for increasing rotation angle. Hence, AFSMR shows a greater robustness against varying transform conditions than FSMR and FSMR without key points. In addition, the proposed AFSMR is nearly $6$~dB better than Lanczos interpolation for a rotation angle of 40 degrees. Furthermore, AFSMR can improve the results for bicubic interpolation in the case of \textit{Zoom 15\%} by more than $10$~dB and a gain of $1.2$~dB is achieved compared to FSMR. In this case, the disregard of key points leads to a worse result than FSMR. Hence, the incorporation of the proposed spectral weighting function is crucial for an improvement in quality. The evaluation in terms of SSIM shows the same behavior. A visual example for the case of \textit{Zoom 15\%} is given in Fig.~\ref{Fig:Results_Visual}. The interpolation based methods in Fig.~\ref{Fig:Results_Visual_Lin}, \ref{Fig:Results_Visual_Cub} and \ref{Fig:Results_Visual_Lanczos} yield blurred results especially for the cables. The same can be observed for the kernel regression in Fig.~\ref{Fig:Results_Visual_Kernel}. As could already expected from the results before, the MLS approach fails also in Fig.~\ref{Fig:Results_Visual_Bose}. Using FSMR produces ringing artifacts particularly on the right side of the right cable in Fig.~\ref{Fig:Results_Visual_Old}. The cable itself is of high quality while no blurring is visible. The proposed AFSMR maintains the high reconstruction quality of the cable but reduces the ringing artifacts next to the cable due to the introduced spectral weighting significantly as shown in Fig.\ref{Fig:Results_Visual_New}. 
\subsubsection{Run Time}
\begin{table}[t]
	\caption{{\label{Tab:ResultsTime}Average processing time in milliseconds per $8\times 8$ block and speed-up factor indicated relative to FSMR.}}
	\centering
	\begin{tabular}{|l|c|c|}
		\hline
		& Time in ms & Speed-up factor\\
		\hline
		Linear & 0.6 &  961.9\\
		\hline
		Cubic & 0.6 &   961.9\\
		\hline
		Lanczos & 15.9 & 34.9 \\
		\hline
		Kernel Reg. & 24.0 & 23.2 \\
		\hline
		MLS & 107.5 & 5.2 \\
		\hline
		FSMR & 555.8 & --- \\
		\hline
		AFSMR & 38.3 & 14.5\\
		\hline		
	\end{tabular}
	\vspace{-.3cm}
\end{table}
The run time measurements are conducted for the application of $15\%$ zoom-in. It is computed on a mid-range computer using the \textit{Intel Xeon(R) CPU E3-1275} with $3.8 GHz$. The results are shown in Tab. \ref{Tab:ResultsTime}. Firstly, the absolute average processing time is given per block and secondly, the speed-up factor with respect to FSMR is indicated. Obviously, the frequency-based methods cannot compete with run times of linear or cubic interpolation, as the used test setup is a non-optimized Matlab implementation. Nevertheless, considering the two high quality model-based resampling methods, AFSMR is 14.5 times faster than FSMR. Moreover, the presented AFSMR can also be reformulated according to \cite{2017_Genser_FastFSR}, resulting in an additional significant speed-up. A further acceleration of the processing time can be achieved if the signal characteristics are taken into account \cite{2018_Genser_PCS}.
\section{Conclusion}
\label{sec:conclusion}
Resampling from arbitrarily located mesh positions onto regularly spaced grid positions is a common task in image processing. One can see this problem on the one hand as classical interpolation problem so that commonly used interpolation methods can be taken for solving it, like e.g. bicubic interpolation. On the other hand, the problem can be assumed to be a reconstruction problem so that more sophisticated solutions like e.g. kernel regression can be applied. An approach that combines both views is FSMR that yields high-quality images. Our proposed AFSMR method is conceptually simpler, yields even better results in terms of PSNR and SSIM, and is at the same time significantly faster than the underlying FSMR. PSNR gains of up to $1.2$~dB can be achieved while the processing time is decreased by a factor of 14.5. These benefits are achieved by disregarding the estimated key points for the model generation and introducing a spectral weighting function at the same time. 
\section{Acknowledgment}
The authors gratefully acknowledge that this work has been supported by the Deutsche Forschungsgemeinschaft (DFG) under contract number KA 926/8-1.
\bibliographystyle{IEEEbib}
\bibliography{2020_MMSP.bib}
\end{document}